\documentclass[letter]{jpsj2}
\usepackage{epsf}
\usepackage{graphicx}
\title{
Quantum Critical ``Opalescence" around Metal-Insulator Transitions}
\author{ Takahiro {\sc Misawa}, Youhei {\sc Yamaji} and Masatoshi {\sc Imada}%
}
\inst{Department of Applied Physics, University of Tokyo, 7-3-1 Hongo, Bunkyo-ku, Tokyo, 113-8656, Japan}

\recdate{April 11, 2006}
\abst{%
Divergent carrier-density fluctuations equivalent to the critical opalescence of gas-liquid transition emerge around 
a metal-insulator critical point at a finite temperature.  In contrast to the gas-liquid transitions, however, 
the critical temperatures can be lowered to zero, which offers a challenging quantum phase transition. We present a 
$microscopic$ description of such quantum critical phenomena in two dimensions. The conventional scheme of phase transitions 
by Ginzburg, Landau and Wilson is violated because of its topological nature.  It offers a clear insight into 
the criticalities of metal-insulator transitions (MIT) associated with Mott or charge-order transitions. Fermi degeneracy 
involving the diverging density fluctuations generates emergent phenomena near the endpoint of the first-order MIT 
and must shed new light on remarkable phenomena found in correlated metals such as unconventional cuprate superconductors. 
It indeed accounts for the otherwise puzzling criticality of the Mott transition recently discovered in an organic conductor. 
We propose to accurately measure enhanced dielectric fluctuations at small wave numbers. }
\kword{metal-insulator transition, quantum critical phenomena, non-Ginzburg-Landau-Wilson transition, Mott transition}
\begin{document}
\maketitle
%{\bf}
%\section{Introduction}
  
Insulating phases appear for various reasons\cite{Mott,RMP,Anderson}, ranging from simple band insulators, 
Anderson insulators caused by randomness, to insulators driven by Coulomb repulsion of electrons such as Mott 
insulators.
The band and Anderson insulators always emerge through continuous MITs from metals at temperature
 $T=0$, 
meaning the zero critical temperature $T_c=0$\cite{Anderson}. However, Mott transitions and charge-order transitions 
often show first-order character with huge jumps in conductivity~\cite{RMP}. The jump decreases 
with increasing temperature and terminates at the critical point at $T=T_c>0$. 
In this letter, we explore the mechanism for the emergence of such first-order and continuous MITs.
% and its consequences. 
%At the border between the first-order and continuous metal-insulator transitions, we find theoretically an unconventional quantum criticality indicating breakdown of conventional Ginzburg-Landau-Wilson scheme. Diverging density fluctuations coexisting with Fermi degeneracy may signal various instabilities and competing orders. 

In the simplest scheme by Ginzburg, Landau and Wilson (GLW) for phase transitions\cite{Landau,Wilson}, the free energy $F$ on the mean-field level is expressed as a function of the order parameter $m$ as  
\begin{equation}
F(m) = \frac{1}{2}A_{GL}m^2 + \frac{1}{4}B_{GL}m^4 - 
mh, \label{eqn:1}
\end{equation}
where we omit the spatial dependence.
Here, $B_{GL}$ is regarded as a positive constant, whereas a phase transition occurs through the sign change of $A_{GL}$, which depends on 
the control parameter $g$ around the critical point $g_c$ as $A_{GL}=A_{GL0}(g-g_c)$. 
Within the mean-field theory, 
when the external field changes the sign from $h>0$ to $h<0$, the free-energy minimum point jumps at $h=0$ from $m_{+}=m_0$ to $m_{-}=-m_0$ 
for $g<g_c$, which determines a critical exponent $\beta$ defined by $\Delta m\equiv |m_{+}-m_{-}|\propto |g-g_c|^{\beta}$ as $\beta=1/2$. 
When $g$ is kept at $g_c$ and $h$ is controlled, another exponent $\delta$ defined by $m\propto h^{1/\delta}$ becomes $\delta=3$. 
The susceptibility $\chi=\lim_{h\rightarrow 0}m/h\propto |g-g_c|^{-\gamma}$ determines the third critical exponent $\gamma=1$.
These mean-field values are modified under fluctuation effects while the mean-field exponents are correct above the upper critical dimension 
$d_c=4$. 
This Ising universality class also applies to the classical gas-liquid transition. The order parameter of the gas-liquid transitions is 
regarded as the particle density measured from the critical point. 

A recent study of (V$_{1-x}$Cr$_x$)$_2$O$_3$ by Limelette {\it et al.}\cite{Limelette} has suggested that the metal-to-Mott-insulator transition also belongs to the Ising universality\cite{Castellani}.  Similarly to the gas-liquid 
transition, the order parameter is interpreted as the carrier density $X$, which drives the singularity of conductivity. 
In analogy with the critical opalescence in the gas-liquid transition, diverging carrier density fluctuations  
control physics around the critical point. 

However, in earlier studies, a scaling theory of a different universality class has been proposed when $T_c$ can be suppressed to 
zero,\cite{Imada1995,Furukawa} in which the simple GLW scheme is violated and unusual exponents $\delta=2$, $\gamma=1$ and $\beta=1$ 
are derived.\cite{Imada2005} A recent experiment\cite{Kanoda} on an organic conductor, $\kappa$-(ET)$_2$Cu(Cn)$_2$Cl, has supported 
the existence of this unconventional universality.

In this letter, we  show that such a new scheme of quantum critical phenomena can indeed be constructed from a {\it microscopic} model.
When the Ising-type critical fluctuations become involved in the Fermi degeneracy region, it deeply modifies the classical nature and a
different type of diverging density fluctuations shows up with emerging effects, indicating the breakdown of the  GLW scheme.   
An unforeseen region of marginally quantum critical (MQC) behavior sandwiched by the regions of the $T_c=0$ critical line and the first-order 
transitions  may signal various instabilities and competing orders.

To understand the essence, the Hubbard model on a $N_s=L\times L$ square lattice defined by
\begin{equation}
H=\sum_{k\sigma}\left( \xi_1 + \xi_2-\mu\right)c_{k\sigma}^{\dagger}c_{k\sigma} + U\sum_i n_{i\uparrow}n_{i\downarrow} + V\sum_{\langle ij\rangle}n_in_j \label{eqn:2}
\end{equation}
is employed with the hopping terms for the nearest neighbor $\xi_1=-2t(\cos{k_x}+\cos{k_y})$ and the next-nearest neighbor 
$\xi_2=4t'\cos{k_x}\cos_{k_y}$. We assume $0<t'<t$.
The standard notation is employed
%Electrons at the site $i$ with the spin $\sigma$ is created (annihilated) by $c_{i\sigma}^{\dagger} (c_{i\sigma})$ and 
their Fourier transforms are denoted by $c_{k\sigma}^{\dagger} (c_{k\sigma})$.  The number operator is defined by 
$n_i=\sum_{\sigma}n_{i\sigma}$ and $n_{i\sigma}=c^{\dagger}_{i.\sigma}c_{i.\sigma}$ 
and the chemical potential is $\mu$.
The onsite and nearest-neighbour Coulomb repulsions are denoted by $U$ and $V$, respectively. 
Hereafter, we take the energy unit as $t=1$. 

We first consider the electron density at half filling, $n =1$. 
When $8V>U$, this model has a tendency of charge ordering, where electrons become alternatingly empty and doubly 
occupied at every other site. 
For $U>8V$, one electron tends to occupy each site leading to the Mott insulator and the exchange interaction 
stabilizes the antiferromagnetic order. 
A dimensionless coupling $g=(8V-U)/2t$ for $8V>U$ and $g=U/2t$ for $U>8V$ is introduced in the following discussion.  

When one assumes the antiferromagnetic or charge orders, the Hartree-Fock approximation (HFA) in the thermodynamic limit 
$N_s\rightarrow \infty$ yields the free energy at temperature $T$ represented by the order parameter $\Delta$ as 
\begin{equation}
F(\Delta)=-\frac{T}{N_s}\sum_{k,\pm} \log\left(1+e^{- E_{\pm}(k)/T}\right) +\frac{\Delta^2}{2g}+\mu n \label{eqn:3}
\end{equation}
with the quasiparticle dispersion
\begin{equation}
E_{\pm}(k) = \xi_2(k) \pm \sqrt{\xi_1(k)^2+\Delta^2} -\mu n, \label{eqn:A4}
\end{equation}
where we take the upper (lower) sign for the wavenumber satisfying $\xi_1>0(<0)$, respectively.  Here the order parameter 
$\Delta$ is defined by $\Delta=gm$ with 
%\begin{equation}
$m=\frac{1}{N_s}\sum_i(n_{i\uparrow}-n_{i\downarrow})e^{iQr_i}$
%\label{eqn:5}
%\end{equation}
for the antiferromagnetic order in the region $U>8V$ and 
%\begin{equation}
$m=\frac{1}{N_s}\sum_i(n_i-n)e^{iQr_i}$
%\label{eqn:6}
%\end{equation}
for the charge order ($U<8V$), with the period being $Q=(\pi,\pi)$.  

HFA ignores fluctuations and thus one might have suspicions on the reliability in strongly correlated systems.  
%Of course, the fluctuations will modify quantitative aspects and the approximation generates some artifact. 
However, as we will see, it clarifies an unexpectedly remarkable feature of MIT overlooked 
in the literature and enables us to provide a microscopic 
description of unconventional non-GLW-type criticality.  
We will also discuss an estimate of the fluctuations beyond HFA later.  

In HFA, MIT takes place in the ordered phase, where $\Delta \neq 0$.  $\Delta$ plays a role of opening a gap, 
but the density of states at the Fermi level remains nonzero when $\Delta \le \Delta_c=2t'$, resulting in a metal. 
This is because the bottom of the upper band 
$E_+$ at  $(\pi,0)$ is lower than the top of the lower band $E_-$ at $(\pi/2,\pi/2)$.  
When $\Delta$ increases with increasing $g$, 
MIT occurs at $\Delta=2t'$, above which the Fermi level lies in the gap, resulting in an insulator.  
Note that the symmetry does not change at MIT.

We first consider $T=0$.  When  the free energy (\ref{eqn:3}) is expanded in terms of $\Delta$ measured from
the transition point $\Delta=\Delta_c$, namely by   $Y\equiv \Delta_c-\Delta$, we obtain
\begin{equation}
F(Y) = F(Y=0)+AY + \frac{1}{2}BY^2 + 
\frac{1}{3!}CY^3+\cdots. \label{eqn:7}
\end{equation}
%The equilibrium is realized at the free-energy minimum and if that 
If the minimum of eq.~(\ref{eqn:7}) is in the region $Y<0$, it describes an insulator, while a metal is given when $Y>0$.

By using the density of states ${\tilde D}_{-}(E)$ (${\tilde D}_+(E)$) for the lower (upper) band, 
the coefficients in Eq.~(\ref{eqn:7}) are given by elementary expansion of Eq.~(\ref{eqn:3}) as  
%\begin{eqnarray}
%A & = & -\frac{\partial F}{\partial\Delta}|_{Y=0} \simeq - \left(\frac{1}{g} - W_1\right)\Delta_c \label{eqn:A24}\\
%B & = & \frac{\partial^2F}{\partial\Delta^2}|_{Y=0} \simeq - \left\{(1+\alpha)Q_++ (1-\alpha)Q_{-}\right\}- W_2+\frac{1}{g} \label{eqn:A25} \\
%C & = & -\frac{\partial^3F}{\partial\Delta^3}|_{Y=0} \nonumber \\
%&\simeq& -3W_3 + \frac{1}{\Delta_c}\left\{(1+\alpha)(3+\alpha)Q'_+ +(1-\alpha)(3-\alpha) Q'_{-} \right\}\nonumber \\
%& &- (1+\alpha)^2R_+ + (1-\alpha)^2R_{-}
%\label{eqn:A25-2} 
%\end{eqnarray}
$A \simeq - \left(1/g - W_1\right)\Delta_c$, 
$B \simeq - \left\{(1+\alpha)Q_++ (1-\alpha)Q_{-}\right\}
- W_2+1/g$
and
$C \simeq -3W_3 + \left\{(1+\alpha)(3+\alpha)Q'_+ +(1-\alpha)(3-\alpha) Q'_{-} \right\}/\Delta_c-(1+\alpha)^2R_++(1-\alpha)^2R_{-}$, where
%begin{eqnarray}
$W_1  \equiv  v_1 $, 
%\label{eqn:A21} \\
$W_2 \equiv  v_1-\Delta_c^2v_3$, 
%label{eqn:A22} \\
and $W_3  \equiv  \Delta_cv_3 - \Delta_c^3 v_5$
%\label{eqn:A23}
%end{eqnarray}
with 
\begin{eqnarray}
v_n \equiv\int_{-\Lambda}^{\Lambda}dE f(E)
\left( \tilde{D}_{-}(E) -\tilde{D}_+(E) \right)\frac{1}{(\xi_1^2+\Delta_c^2)^{n/2}}.
\label{eqn:A23-2}
\end{eqnarray}
%\begin{eqnarray}
%Q_s & = & \int_{-\Lambda}^{\Lambda} dE
%\phi(E,T)\tilde{D}_s(E), \label{eqn:A25-3} \\
%Q'_s & = & \int_{-\Lambda}^{\Lambda} dE
%\phi(E,T)\tilde{D}_s(E)^2/\tilde{D}^{(0)}_s(E), \label{eqn:A25-4} \\
%R_s & = & \int_{-\Lambda}^{\Lambda} dE
%\phi(E,T)\frac{\partial \tilde{D}_s(E)}{\partial E}, \label{eqn:A25-5} \end{eqnarray}
We have also introduced $Q_s=\int_{-\Lambda}^{\Lambda} dE \phi(E,T)\tilde{D}_s(E)$,
$Q'_s=\int_{-\Lambda}^{\Lambda} dE \phi(E,T)\tilde{D}_s(E)^2/\tilde{D}^{(0)}_s(E)$, and
$R_s=\int_{-\Lambda}^{\Lambda} dE
\phi(E,T)\partial \tilde{D}_s(E)/\partial E$, for $s=+$ and $-$ with the band cutoff $\Lambda$. The density of states for $t'=0$ is 
denoted by $\tilde{D}^{(0)}_s(E)$. 
Because the energy derivative of the Fermi distribution $f$ defined by 
%\begin{eqnarray}
$\phi(E,T) \equiv -df(E)/dE=e^{E/T}/[T\left(e^{E/T} +1\right)^2]$
%\label{eqn:A19-2}
%\end{eqnarray}
%$\phi(E,T) \equiv  df(E)/dE=e^{E/T}/T\left(e^{E/T} +1\right)^2$
is reduced to the delta function $\delta(E)$ in the limit $T\rightarrow 0$, 
$Q_s$ and $Q'_s$ reduce to $Q_s=\tilde{D}_s(0)$ and ${Q'}_s=\tilde{D}_s(0)^2/\tilde{D}^{(0)}_s(0)$, respectively, at 
$T=0$. Then $Q_+$ and $Q_{-}$ as well as $Q_+'$ and $Q_{-}'$ jump from the metallic to insulating sides, since $\tilde{D}_+(E)$ 
and $\tilde{D}_{-}(E)$ for two-dimensional systems jump from nonzero to zero at the Fermi level $E=0$.
The factor
$ \alpha \equiv -\partial \mu/\partial\Delta $
is defined by the relation 
%between the chemical potential $\mu$ and $Y$ as 
$\mu=\alpha Y$ with 
%\begin{equation}
$\alpha=-(\sqrt{t^2-{t'}^2}-2 t')/(\sqrt{t^2-{t'}^2}+2 t')$.
%\label{eqn:A36}
%\end{equation}
The carrier density for $Y>0$ is easily derived as
\begin{eqnarray}
X & = & 2\frac{\pi}{\Delta_c}(1+\alpha)Y, \label{eqn:A37}
\end{eqnarray}
which holds for $t'>0$.
%If the coefficients $A,B$ and $C$ are regarded as constants, we realize that Eq.~(\ref{eqn:7}) does not follow the simplest GLW form of Eq.~(\ref{eqn:1}).
%This is because the cubic term absent in Eq.~(\ref{eqn:1}) by symmetry reason is present.  

\begin{figure}[h!]
\begin{center}
\includegraphics[width=6cm]{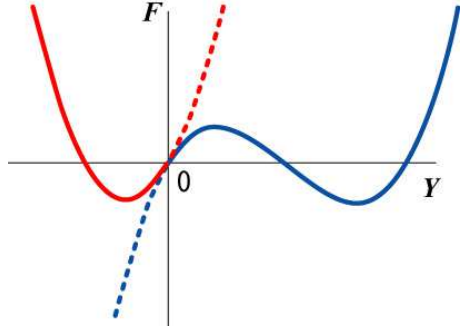}
\end{center}
\caption{(color): Free energy of extended Hubbard model as a function of 
%the gap measured from the metal-insulator transition point, 
$Y=\Delta_c-\Delta$. The red (blue) cubic curve representing the insulating region $Y<0$ (metallic region $Y>0$) 
in total form a double-well structure as a combined solid curve.}
\label{two-cubic}
\end{figure}

Now it turns out that the coefficients $B$ and $C$ explicitly depend on the density of states at the Fermi level, which is nonzero in 
the metal, while it jumps to vanish in the insulator.  Therefore, $B$ and $C$ jump at the MIT point at $T=0$ 
between $B_m$ and $C_m$ in the metal ($Y>0$) and $B_i$ and $C_i$ in the insulator ($Y<0$), respectively. 
We realize that such jumps of 
the coefficients violate the GLW scheme of conventional expansion.
When we look into more details, $C_i<0<C_m$ and $B_m<B_i$ are satisfied for $0<t'<t$. Furthermore, within the realistic choice of 
parameters, a minimum appears at $Y=Y_i\sim -A/B_i<0$ if $B_i^2 \gg AC_i>0$ and another minimum appears at 
\begin{equation}
Y=Y_m\sim (-B_m+\sqrt{B_m^2-2AC_m})/C_m>0
\label{ym}
\end{equation}
if $B_m<0$ and $B_m^2-2AC_m>0$.
An example of such free energy is illustrated in Fig. \ref{two-cubic}, where the overall form is 
given by connecting two cubic functions generating a double-well structure in total.  
The first-order MIT occurs when the minimum changes between $Y_i$ and $Y_m$.
%%%%%%%%%%%%%%%%%
% Fig.1
%%%%%%%%%%%%%%%%

\begin{figure}[h!]
\begin{center}
\includegraphics[width=10cm]{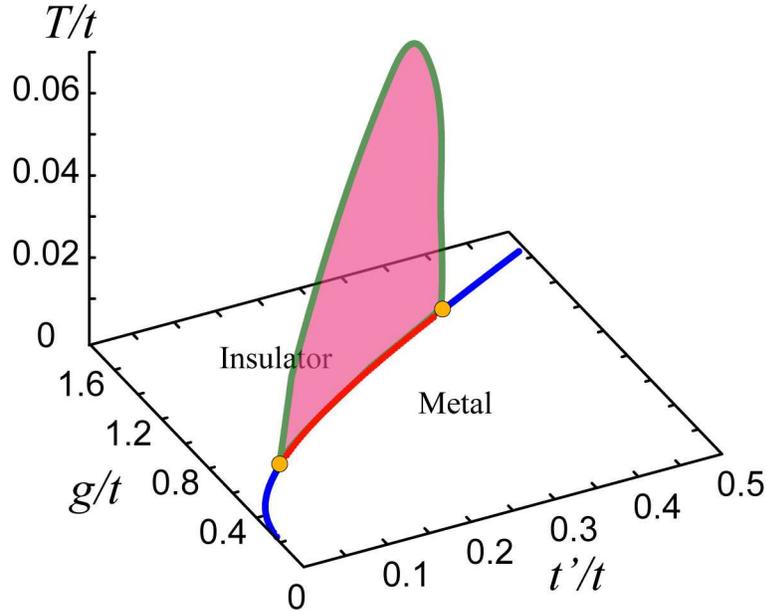}
\end{center}
\caption{Phase diagram of metal and insulator in parameter space of temperature $T$, $g$ and $t'$ for Hubbard model 
on square lattice. 
The metal-insulator boundary set by the first-order MIT is illustrated by the pink surface surrounded by critical line shown by green curve. 
The endpoints of the critical line are marginal quantum critical points (MQCPs) depicted by yellow circles. 
Continuous and first-order MITs at $T=0$ are shown by blue and red curves, respectively.
}
\label{Mi_Phase_5}
\end{figure}

This result indicates that electron correlation effects contain a microscopic mechanism of generating a first-order 
MIT through a synergetic and mutual feedback between the growth of the order parameters and 
MIT at the Fermi level. 
%This is in contrast to metal-insulator transitions of the noninteracting electrons and cannot be described by a simple transition to the band insulator.
Although such a mechanism of the first-order MIT has not been well recognized in the literature\cite{KondoMoriya}, the present analytical 
results for small $Y$ are indeed confirmed in our numerical Hartree-Fock solution of the Hubbard model, where the first-order MIT 
appears for $t_{c2}'/t=0.0557<t'/t<0.3645=t_{c1}'/t$ (see Fig. \ref{Mi_Phase_5}). 
%\begin{figure}[h]
%\begin{center}
%\includegraphics[width=4cm]{Phase_ZT_5.eps}
%\end{center}
%\caption{Phase diagram of metal and insulator at $T=0$ in the parameter space of coupling constant $g$ and the next-nearest neighbour transfer $t'$ for the Hubbard model obtained by numerically solving the Hartee-Fock equation. First-order transition of metal-insulator boundary is illustrated by red and pink curves. The red plot with diamonds is for the transition between paramagnetic metal (PM) and ordered insulator (OI), while the pink one with crosses is between OI and ordered metal (OM). OI and OM have either charge order or antiferromagnetic order, depending on the sign of $U/2-4V$ (see text). Blue curve with squares shows continuous transition between OM and OI. Marginal quantum critical points (MQCP) are plotted as yellow circles, which appear always between OM and OI.}
%\label{Phase_ZT_5}
%\end{figure}

With elevating temperatures, the first-order jump in the region $t_{c2}'<t'<t_{c1}'$ vanishes at $T_c$, 
which forms the critical line illustrated as a green curve in Fig.\ref{Mi_Phase_5}. The critical line is given by the 
crossing line of 
the $A=0$ surface and the $B_m=0$ surface in the parameter space of $T$, $g$ and $t'/t$, where the left and right 
minima in Fig.\ref{two-cubic} 
merge at $Y=0$. 
%%%%%%%%%%%%%%%%%%%%%%%%%%
% Fig.2
%%%%%%%%%%%%%%%%%%%%%%%%%%%%
%We note that the transition becomes continuous if it occurs through continuous change of $A$ from $A>0$ to $A<0$ even when $B$ and $C$ jump at $Y=0$ because the first derivative of the free energy $\partial F/\partial Y$ at the transion point $Y=0$ becomes continuous. 
On the other hand, in the region $t'>t'_{c1}$ or $t'<t'_{c2}$, $B_m$ becomes positive, where the first-order MIT does not exist any 
more, while the continuous MIT boundary further continues at $T=0$ as blue curves, which is given from the 
crossing line of two surfaces, $A=0$ and $T=0$. 
At $T=0$, the boundaries between the quantum critical lines (blue curves) and the first-order region (pink surface) appear at $t_{c1}'$ 
and $t_{c2}'$, which are illustrated as yellow circles in Fig.\ref{Mi_Phase_5}. We call these two points the marginally quantum critical 
points (MQCPs), because they appear at the border (or margin) from the quantum critical line to the finite-temperature classical critical 
point.  In other words, each MQCP is determined from the cross point of all the three surfaces $A=0$, $B_m=0$ and $T=0$.  It is remarkable 
that two MQCPs appear and $T_c$ forms a dome structure sandwiched by these MQCPs.
%Across this boundary, the free-energy minimum changes continuously through $Y=0$.  
The overall phase diagram illustrated in Fig.~\ref{Mi_Phase_5} proves that a simple two-dimensional Hubbard model allows us to study 
the diversity of MITs varying from (i) the first-order, (ii) marginally quantum to 
(iii) continuous (quantum critical) ones at $T=0$~\cite{details}.

%Quantum criticality is a field of recent extensive studies\cite{Sachdev}.  For itinerant electronic models, it appears at the endpoint of the magnetic transition, where the critical temperature is lowered to zero beyond which the transition disappears~\cite{Moriya,Hertz,Millis}. The quantum critical endpoint of the first order transition surface has also been discussed in the context of metamagnetic transition within metallic phases\cite{Millis2}.  In our case, however, a crucial difference from them is that the quantum critical line continues as the blue curves in Fig.~\ref{Mi_Phase_5}, because metals and insulators must always have clear distinction at $T=0$\cite{Imada2005}. 

\begin{figure}[h!]
\begin{center}
\includegraphics[width=10cm]{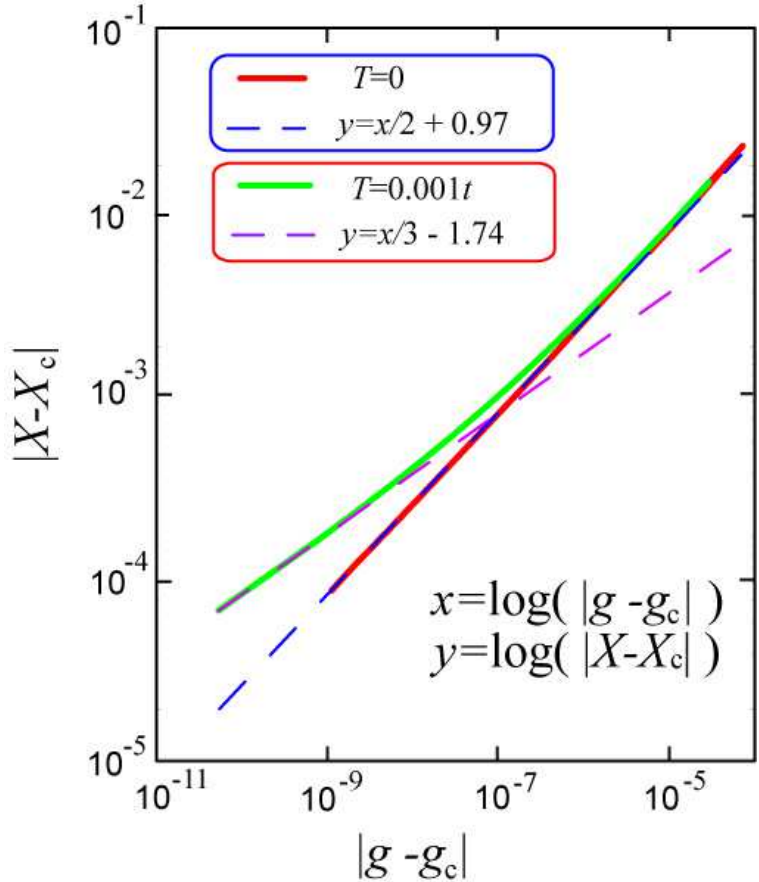}
\end{center}
\caption{Scaling of carrier density measured from critical point $|X-X_c|$ as functions of $g-g_c$ near MQCP at $t'/t=0.3645$, 
which determines exponent $\delta$. At $T$=0, MQC scaling, $\delta=2$, is always satisfied, while a crossover from 
$\delta=2$ to $\delta=3$ is seen at finite temperatures. Results are obtained by numerically solving the Hartree-Fock equations around MQCP.}
\label{Carrier_Cross}
\end{figure}

Around MQCP, we may expand as $B_m =B_m^{(g)}(g-g_c)$ or $B_m=B_m^{(T)}(T-T_c)$ with $T_c=0$, where $B_m^{(g)}$ and $B_m^{(T)}$ 
are constants. 
Because of Eqs.~(\ref{eqn:A37}) and (\ref{ym}), the jump of $X$ at the first-order MIT grows as 
$X \propto Y_m \propto 2B_m^{(g)}(g-g_c)^{\beta}/C_m$, with an unusual critical exponent $\beta=1$.   
When a control parameter $h$ such as pressure couples linearly to $Y$ in the free energy, namely, if $-Yh$ is added to the free energy 
at MQCP given by $A=B_m=0$, the free energy minimum occurs at $C_mY^{\delta}/2-h=0$ on the metallic side, leading to 
$X\propto h^{1/\delta}$ with the exponent $\delta=2$. Figure \ref{Carrier_Cross} shows $\delta=2$ at $T=0$ in the growth of 
carrier number as a function of $g$ measured from MQCP at $g=g_c$, where $g-g_c$ plays the role of $h$ here.
%On the other hand, in the insulating region, the Hartree-Fock theory predicts the opening of the insulating gap $h \sim B_i Y+(C/2)Y^2$.  Since $B_i$ may remain nonzero in general, the gap opens linearly with $h$ resulting in $\delta_i=1$. 
When one approaches MQCP strictly along $A=0$, the susceptibility scales as $\chi \equiv dY/dh \propto 1/B_m \propto |g-g_c|^{-\gamma}$ with $\gamma=1$. These show that the carrier density fluctuations diverge at MQCP either as $\chi \propto |g-g_c|^{-1}$ or as $\chi \propto X^{-1}$.
We recall that the critical opalescence signals diverging compressibility in liquid-gas transitions. Such divergence also emerges in an electronic system even at $T=0$, but with different exponents. 
All the conventional scaling relations such as $\beta(\delta-1)=\gamma$ and the hyperscaling relation are still satisfied.

The unusual exponents $\delta=2$, $\beta=1$ and $\gamma=1$ at MQCP are completely consistent with the scaling theory\cite{Imada1995} 
and the quantum Monte Carlo results for the filling-control MIT.\cite{Furukawa}
Moreover they are indeed in agreement with recent experimental results obtained for a quasi-two-dimensional $\kappa$-ET-type organic conductor\cite{Kanoda}.   

\begin{figure}[h!!]
\begin{center}
\includegraphics[width=12cm]{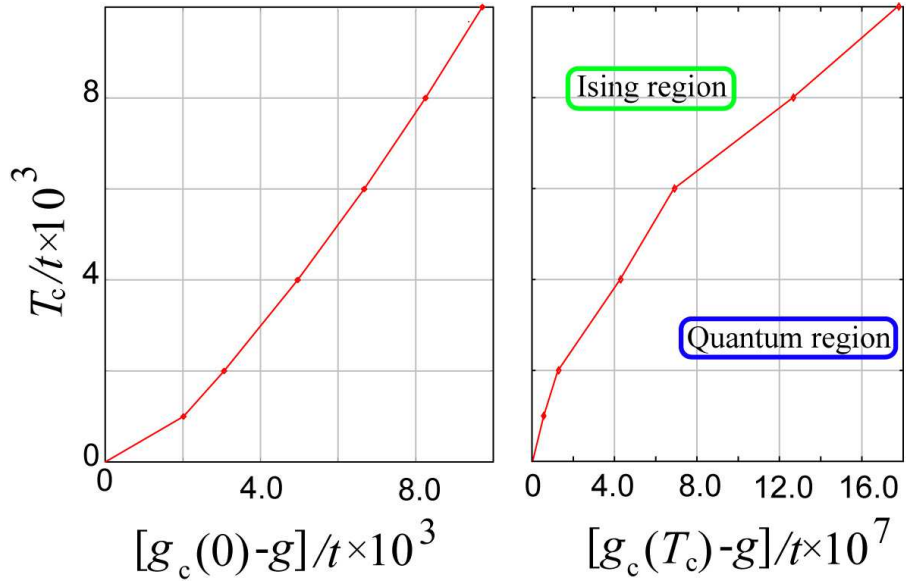}
\end{center}
\caption{(color): Left panel: Critical temperature plotted against $g(T_c=0)-g$ at $t'/t=0.3645$. Right panel: 
Ising and MQC regions separated by crossover. The crossover point is defined by the crossing point of two 
asymptotic lines in the plane of $X-X_c$ and $|g-g_c|$ as in an example of two dashed lines in Fig.~\ref{Carrier_Cross}. 
%asymptotic lines as seen in an example of two dashed lines in Fig.4. 
The distance between the critical point and the crossover measured by $|g-g_c(T_c)|$ (abscissa) is plotted as a function of $T_c$ (ordinate). 
%For $T_c=0.01t$, the crossover occurs at $|g-g_c|/t\sim 1.8\times 10^{-6}$.
}
\label{crossover}
\end{figure}

When $T_c$ becomes nonzero away from MQCP, we expect the emergence of a crossover between the marginal quantum criticality and 
the conventional Ising criticalities. 
%This crossover can also be studied in the present approach.  
At $T>0$, the jumps in $B$ and $C$ immediately smear out leading to effective $Y$ dependences of $B$ and $C$ and 
the GLW expansion (\ref{eqn:1}) ultimately becomes valid.  However, at sufficiently low temperatures, the change from 
$B_i$ to $B_m$, as well as from $C_i$ to $C_m$, occurs within a very small interval, $|Y|<T$. It turns out that if $|Y|>T$ 
is satisfied, the marginal quantum criticality dominates, while the crossover to the Ising criticality occurs around $|Y|\sim T$. 
%In other words, the crossover occurs around $T\sim(B_m^{(g)}/C_m)(g-g_c)$ or $T\sim (B_m^{(T)}/C_m)|T-T_c|$.
This crossover at finite temperatures is indeed seen in our numerical solution as seen in Fig.~\ref{Carrier_Cross}.
The left panel of Fig.~\ref{crossover} illustrates how $T_c$ grows as a function of $g_c(T=0)-g$ and in the right panel 
we show how the Ising region with $\delta=3$ around $T_c$ becomes extended with increasing $T_c$ while most of the region 
is governed by MQCP with $\delta=2$. The Ising region is very limited ($|g-g_c(T_c)|/t<10^{-6}$) even for $T_c=0.01t$, 
whereas the critical region governed by MQCP is wide. 

In the organic coductor\cite{Kanoda}, the exponents are measured around a finite-temperature critical point.  
However, the present unconventional criticality at MQCP may indeed be observed, because the experimental resolution is 0.1 MPa, 
while for a realistic choice of parameters and $T_c\sim 0.01t$, the Ising region appears only for the pressure $P$ measured 
from the critical value $|P-P_c|< 10^{-3}$ MPa, as derived from 1 MPa corresponding to change in $0.002t$ in the case of 
$\kappa$-(ET)$_2$[N(CN)$_2$]Cl ~\cite{Mori}. 
Although our analysis using the mean-field approximation of the Hubbard model may oversimplify the complexity, a more than 
two orders magnitude difference certainly supports the idea that the quantum criticality should be seen. 
%More detailed quantitative analyses on finite-temperature effects and relation to the experimental condition are found in Supplementary Discussion for Finite Temperature Effect.

%In this organic conductor, one needs to approach much closer to the critical point to see the Ising universality. 
%For high critical temperatures, the Ising universality becomes wide as is established in (V$_{1-x}$Cr$_x$)$_2$O$_3$ with high $T_c\sim 450$K\cite{Limelette}.  
%So far we have considered the two-dimensional case. However, when the first-order transition exists, we obtain similar phase diagram in other dimensions. 
%%%%%%%%%%%%%%%%%
% Fig.3
%%%%%%%%%%%%%%%%
%%%%%%%%%%%%%%%%%
% Fig.4
%%%%%%%%%%%%%%%%
In its literal form, the present non-GLW picture appears to apply when some order such as antiferromagnetism exists in the both sides 
of the transition.  However, the opening of the preexisting gap itself in metals is not necessarily the consequence of the symmetry 
broken order.  
The quasiparticle gap between the so-called lower and upper Hubbard bands may open without the symmetry breaking.\cite{Hanasaki}
%, as known in the example of the Mott insulator in low dimensions. 
Therefore, the present consequence may survive without the strict order, where $\Delta$ is replaced with a correlation-induced gap 
without long-ranged orders.

%When one calculates the fluctuation beyond HFA, it turns out that it does not grow near MQCP.  
%Therefore, the present universality class is valid. 
%In other words, the Ginzburg criterion shows that the system is always at the upper critical dimension in any dimension and the mean-field theory is justified\cite{Ginzburg}.  See Supplementary Discussion for Ginzburg Criterion for the proof. 

When filling is controlled, the first-order MIT appears as a phase separation\cite{Emery}, and the MQCP appears at the boundary 
of this phase separation region at $T=0$.
Around MQCP, the carrier density is scaled by $X\propto \mu^{1/\delta}$ with the chemical potential $\mu$ and $\delta=2$.  
This results in the charge compressibility $\kappa=dX/d\mu\propto X^{1-\delta}=X^{-1}$. The compressibility also shows 
$\kappa\propto |g-g_c|^{-\gamma}$ with $\gamma=1$ similarly to the case of bandwidth control. Therefore, $\kappa$ generically 
diverges in all the routes towards MQCP.  It should be noted that quantum Monte Carlo studies~\cite{Furukawa} 
have also shown $\kappa \propto 1/X$, implying that the present result holds beyond the mean-field approximation.

The present result may be justified beyond the mean-field approximation for the following reason: 
$X$ has its  intrinsic length scale $\xi=1/\sqrt{X}$ in the metallic side, because the carrier density 
$X$ has the unique diverging length scale of the mean carrier distance, $\xi$, yielding 
$\xi \propto (g-g_c)^{-\nu} \propto B^{-\nu} \propto X^{-\nu}$ with $\nu=1/2$ defined at $A=0$.  
Then the dynamical exponent $z$ of MQCP becomes 4 because $\kappa/X \propto \xi^{z}$ is satisfied~\cite{RMP,Imada2005,details}.  
The mean field approximation is marginally justified, because the Ginzburg criterion\cite{Ginzburg} for the upper critical dimension given by 
$\gamma+2\beta-\nu (d+z)=0$ is satisfied for $\gamma=1$, $\beta=1, \nu=1/2, d=2$ and $z=4$. The critical 
exponents then become exact except for logarithmic corrections. 
%Our finding shows, in a microscopic theory of metal-insulator transitions described by a Hubbard model, that the electron correlation generates a synergy effect to make the metal-insulator transition first-order character.  With increasing temperatures, this first-order transition ends at the critical point similarly to the classical gas-liquid transition.  However, the critical temperature can be lowered to zero in the metal-insulator transition, which yields an unusual behavior called MQC phenomena.  

%At both of the bandwidth and filling control MQCP, the diverging carrier density fluctuations coexist with the Fermi degeneracy. This divergence may be called ``quantum critical opalescence" in analogy with the critical opalescence around the gas-liquid critical point. 
Such density fluctuations of degenerate Fermi particles may induce various instabilities toward orders including superconductivity, 
and profound effects on competing orders may be triggered.
Our findings open a new route of studying the physics of electron correlation effects.
%, where an unusual universality class is identified for metal-insulator transitions beyond the Ginzburg-Landau-Wilson scheme and dramatic carrier density fluctuations occur at MQCP on the boundary of continuous and first-order transitions. 
The result implies the crucial importance of enhanced dielectric fluctuations typically in the energy range up to the insulating gap 
amplitude (typically up to $\sim$ 1eV) at a small wavenumber.  Even in the filling control transition with the constraint of the long-ranged 
Coulomb force ignored in the Hubbard model, such enhanced dielectric response may play a crucial role in the dramatic change of spectral 
properties over a large energy range as known in optical conductivity~\cite{RMP}.  
It is desired to develop experimental probes for carrier density-density correlations beyond the present accuracy of available probes 
such as electron energy loss spectroscopy (EELS) and X-ray. In fact, an accurate probe should show ``opalescence".

{\small {\bf Acknowledgements} 
The authors thank K. Kanoda and F. Kagawa for useful discussions.
A part of this work has been supported by a Grant in Aid for Scientific Research on Priority Area  from the Ministry of Education, Culture Sports, Science and Technology.}

%\end{document}
%%%%%%%%%%%%%%%%%%%%%%%%%%%%%%%%%%%%%%%%%%%%%%%%%%%%%%%
%\vskip -0.5cm
%\newpage
%\vskip -3cm
%\vskip -3cm
%\vskip -3cm
%%%%%%%%%%%%%%%%%%%%%%%%%%%%%%%%%%%%%%%%%%%%%%%%%
\end{document}